\documentclass{elsart}
\usepackage[dvips]{graphicx}
\usepackage{amssymb}

\begin{document}
\begin{flushright}
WSU-HEP-0206
\end{flushright}

\begin{frontmatter}

\title{
Track Timing at $e^{+}e^{-}$ Linear Colliders with a Silicon Drift
Detector Main Tracker
}

\author[]{R.~Bellwied}, 
\author[]{D.~Cinabro}, 
\author[]{V.~L.~Rykov\thanksref{CA}}
\address{Wayne State University, Detroit, MI 48201, USA}
\thanks[CA]{Corresponding author. Phone: 
(313)--577--2781; fax: (313)--577--0711;
e--mail: rykov@physics.wayne.edu}

\begin{abstract}
   The track timing capabilities of a silicon drift detector based
tracker for a future linear electron-positron collider are
evaluated. We show such a detector can time tracks at the nanosecond,
and for high-$P_{T}$, sub-nanosecond level. This implies that, even
for collider designs with the bunch spacing at 1.4~$ns$, every track
can be assigned to a particular bunch crossing at a confidence level
of up to two standard deviations. We suggest a choice for the drift
axes in the tracker layers to simultaneously optimize the momentum
resolution and track timing.
\end{abstract}

\end{frontmatter}

\section{Introduction}
\label{sec:intro}

    Projects for future $e^{+}e^{-}$\, linear colliders (LC) operating
at $\sqrt{S}$ = \mbox{0.5--1} TeV~\cite{nlc:01,jlc:97,tesla:tdr}
consider a number of detector
layouts~\cite{lcphys:01,jlcphys:01}. For example, the NLC
proposal~\cite{lcphys:01} includes two options for the high-energy
interaction region (IR) which are called Large, {\em L}; Silicon
Detector, {\em SD\/}; and one for a low energy second IR called
Precise, {\em P}. The tracking systems in all three layouts include a
high-resolution pixel vertex detector (VXD), but differ significantly
in the technology choices for the main tracker (MT). The central MTs
for the {\em L}- and {\em P}-configurations are based on large-volume
time projection chambers (TPC). In the {\em SD}-version, the central
MT consists of few tracking layers of either silicon drift (SDD), or
silicon strip detectors.

   In all three detector layouts, drift detectors with
three-dimensional space point measurements along charged particle 
trajectories are considered as either the only (TPC), or one of few
alternative solutions (SDD) for the central MT. Drifting of generated
electron clouds in the TPC and SDD is rather slow. This leads to their
long sensitivity to ionization which ranges from a few to a few tens
of microseconds. Therefore, tracks from the event of interest will
coexist in the raw data with tracks from events which occurred at some
time before and after the trigger. These tracks need to be recognized
and separated from the triggered event.

   The time structures of collisions at the various proposed high
energy LCs are expected to be similar, but differ in detail. All three
projects feature trains (Rf-pulses) of $e^{-}$ and $e^{+}$ bunches. At
NLC/JLC, each train will consist of 190 bunches, separated by
1.4~$ns$, resulting in a train duration of $\sim$265~$ns$. TESLA
features 950~$\mu s$\, long trains of $\sim$2800 bunches separated by
337~$ns$. The projected Rf-pulse repetition rates at NLC, JLC and
TESLA are 120, 100 and 5~$Hz$, respectively. The design luminosities
in all three proposals are on the order of
\mbox{$\sim$(2--3)$\cdot$10$^{34}$~$cm^{-2}s^{-1}$}.

   According to the estimates of Ref.~\cite{lcphys:01} for the design
luminosity, $\sim$2.2 hadronic $\gamma\gamma$\, events/train, on
average, will occur at the NLC/JLC in addition to the trigger. The
average number of tracks is $\sim$17 with $\sim$100~$GeV$\, deposit in
the  calorimeter per such an event. All tracks from these events,
along with the trigger, will be present in a TPC or SDD based MT
simultaneously during an ionization drift, which is considerably
longer than the NLC/JLC train time-length\footnote{Due to the low 
Rf-pulse repetition rates, there will be no more than one train per
SDD or TPC drift cycle, though.}. At TESLA, there will be $\sim$0.02 
hadronic $\gamma\gamma$\, events per bunch
crossing~\cite{hensel:00,lcphys:01}, and ionization from only a
fraction of a single train will be present simultaneously in the SDD
or TPC. These transfer into $\sim$0.5 background events occurring
during the SDD drift of \mbox{7--8~$\mu s$}, and $\sim$3--5 events
during the \mbox{50--60~$\mu s$} long TPC drift in {\em L}- and {\em
P}-versions of the detector. In all cases, the beamstrahlung
background from $e^{+}e^{-}$\, bunch crossings will also be present in
the trackers. It is recognized~\cite{lcphys:01} that, if the {\em time
stamping} for the tracks in the TPC or SDD is not done, it could
seriously impact the detector performance, particularly its missing
mass resolution.

   Matching tracks in the TPC or SDD to the collision region and/or
the VXD provides some time stamping. However, this does not work for
secondary tracks from decays of long lived particles. Also the VXD,
located at a small distance to the collision region, has a much higher
occupancy of hits from low-$P_{T}$ tracks produced by beamstrahlung,
making matching more difficult. Therefore, it is more desirable to
find a solution that uses information from only the MT. In the case of
a TPC, it was suggested to place at some TPC depth a fast intermediate
tracker, constructed from scintillating fibers, and alternatively or
in addition, a silicon intermediate tracking detector just inside the
TPC inner radius~\cite{lcphys:01}. A similar solution 
may also be possible for a SDD based tracker. However, due to the
finer segmentation of the SDDs and more flexibility in the choices for
SDD drift axes in MT layers, such special time-stamping layer(s) will
not be necessary if a silicon drift detector is chosen as the
technology for the central MT.

   In this note, we present the estimates for the ``self-timing''
capabilities of the SDD based MT with various choices for drift axes
in its layers. A similar approach was suggested for use in the STAR
experiment at RHIC to resolve the high-luminosity $pp$\, event
pile-up~\cite{sn:169,sn:237}.

\section{{\em SD} main tracker layout and simulation model}

\subsection{SDD based main tracker}
\label{sec:setup}

\begin{figure}[htb]
\centerline{\hbox{
\includegraphics*[width=2.6in,bb=20 150 550 590]{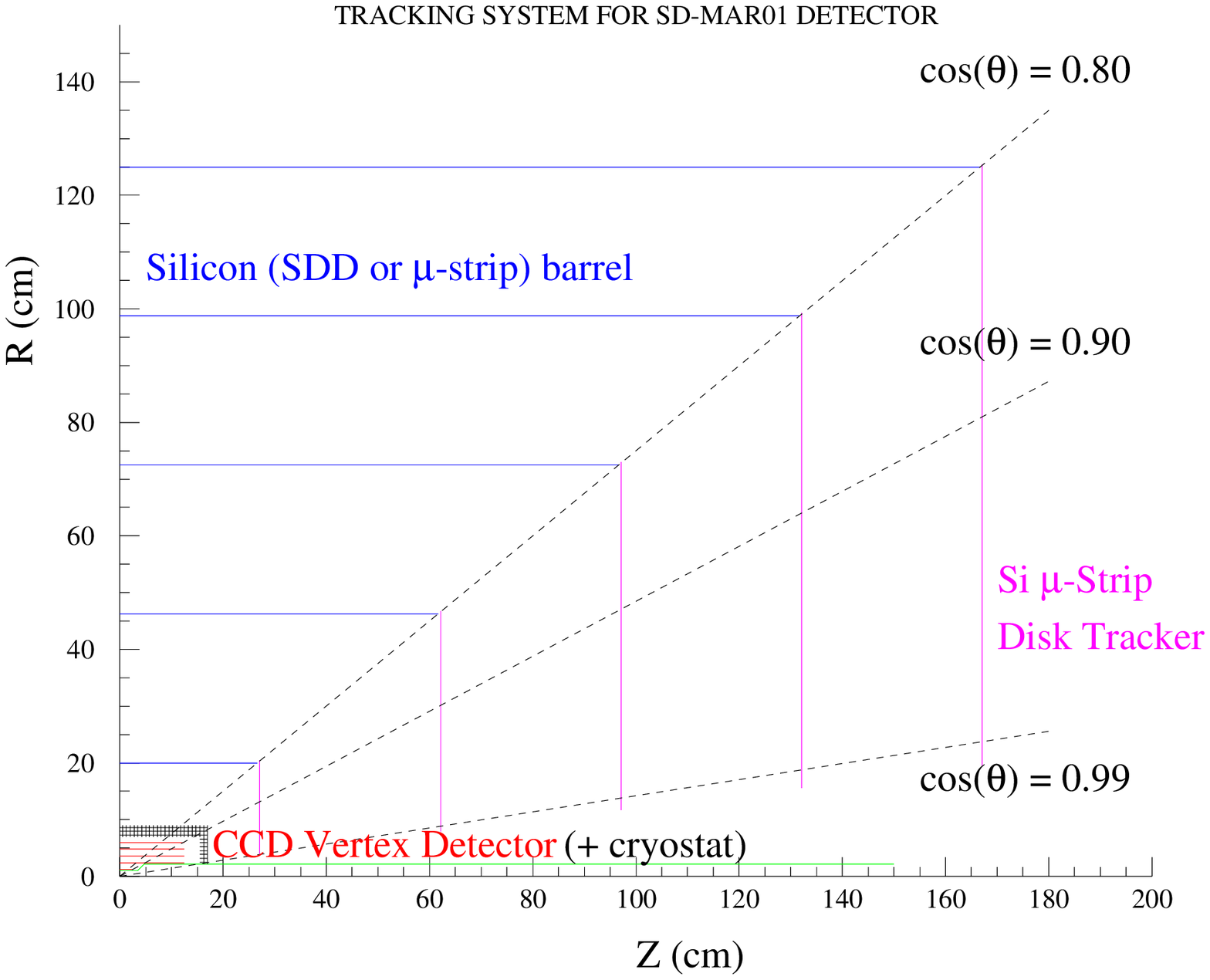}
\includegraphics[width=2.7in,bb=80 20 560 450]{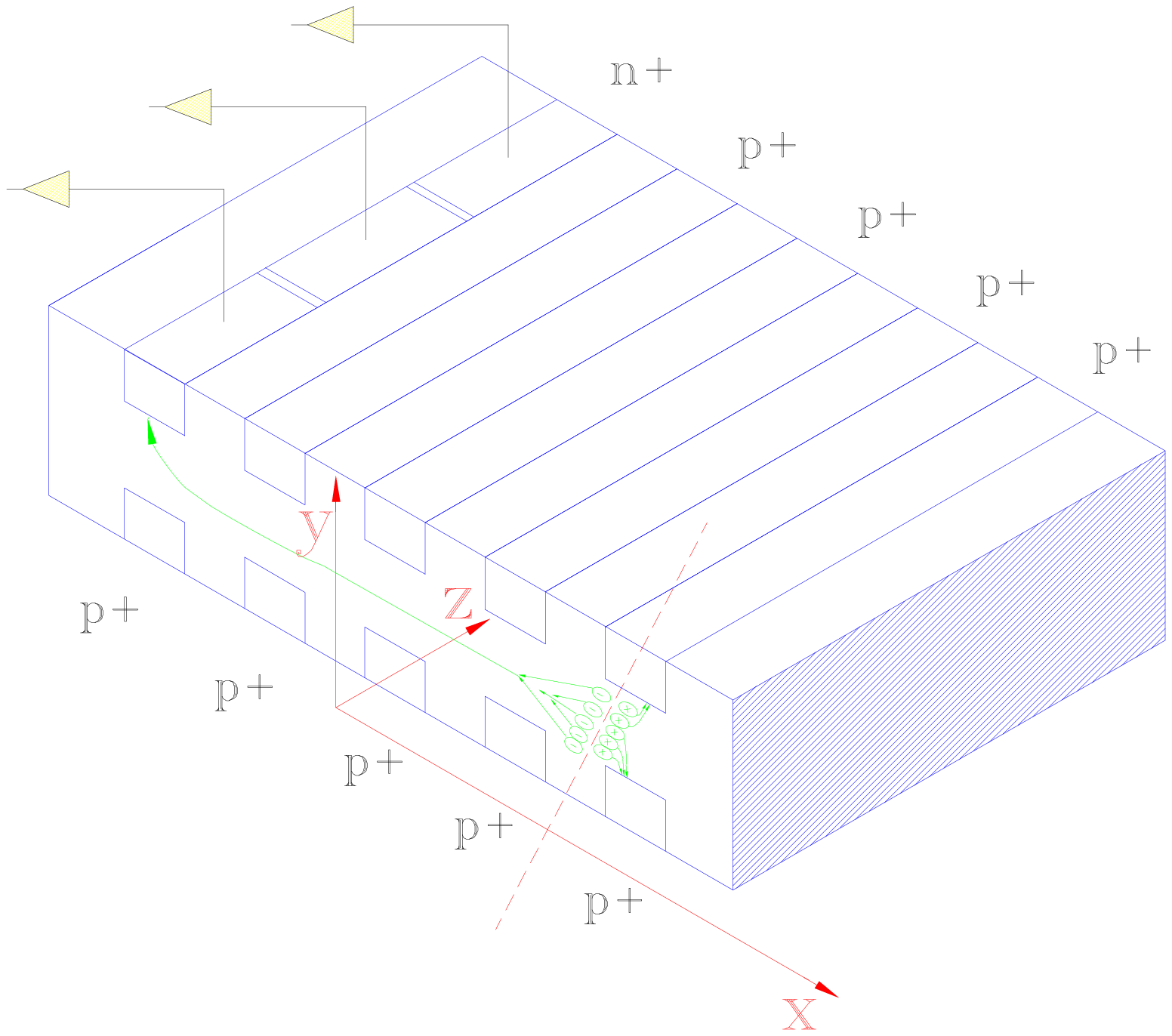}
}}
\caption[]
{
\footnotesize
Sketch of {\em SD\/} tracking system (left) and the SDD schematic view
(right).
\normalsize
}
\label{fig:setup}
\end{figure}

The tracking system of the proposed {\em SD\/}
option~\cite{lcphys:01,schumm:01} with the 5~$T$\, strong magnetic
field is schematically shown in Fig.~\ref{fig:setup}, left. Its
central and forward MTs consist of a 5-layer silicon barrel, made of
either silicon drift or micro-strip detectors, and five layers of
double-sided silicon micro-strip forward disks, respectively. The
layers of the central MT barrel are located at radii from 20 to
125~$cm$. The 5-layer CCD vertex detector is located closer to the
collision region at radii from 1.2 to 6~$cm$.

   The SDD based version of the central MT will be comprised of
$\sim$6000 SDDs of 10$\times$10~$cm^{2}$\, area and thickness
\mbox{150--200~$\mu m$} each. A two-dimensional position of a
particle hit in the SDD wafer is measured as follows~\cite{rehak:90}
(see SDD schematic view in Fig.~\ref{fig:setup}, right). After passage
of an ionizing particle through the SDD, the created electron cloud is
within \mbox{$\sim$5--10 $ns$}\, confined to a \mbox{$\sim$10--20~$\mu
m$}\, thick ``pancake'' around the SDD middle plane. At the same time,
the cloud starts drifting along the $x$-axis\footnote{The reference
frame of Fig.~\ref{fig:setup}, right, is used here and to the end of 
Sec.~\ref{sec:setup}.} in the uniform electric field
\mbox{$E_{x}\simeq$ 40-50 $V/mm$}, applied to the SDD, with a constant
velocity $V_{drift}~\simeq$~6-7~$\mu m/ns$. The electron cloud
eventually drifts to the anodes, located at the SDD edge. The SDD
anodes are spaced every \mbox{100--300 $\mu m$} (for the LC detector,
the exact spacing is still to be optimized). The $x$-position of a
particle crossing is determined by measuring the time delay between
the trigger and anode signals, and the $z$-position is determined from
the signal distribution over hit anodes. An estimate for the
practically achievable spatial resolution of the SDD based tracker
along the anodes ($z$-axis) is \mbox{$\sim$5--7 $\mu m$}. The
resolution in the drift direction ($x$-axis) is expected to be
somewhat worse, at about \mbox{8--10 $\mu m$}, due to drift
nonlinearities caused by defects in material, environmental effects
and calibration uncertainties.

   In order to reduce the maximum voltage on a wafer, previous
designs~\cite{e896:00,svt:00} feature a ``double-SDD'', i.e. wafers
consisting of two SDDs, drifting along the same axis but in opposite
directions. The SDD anodes are located at the two opposite edges of
the wafer, and high voltage, creating the drift field, is applied in
the middle of the wafer at $x = const$. The alteration of the drift
{\em direction} within such small MT pieces, as each single SDD wafer,
represents a powerful tool to discriminate tracks by their generation
time. The other handle for doing this is the alteration of the drift
{\em axes} in the MT layers, which is also evaluated in this paper.

\subsection{Simulation model}

   For the studies presented here, a simplified model of the {\em
SD\/} tracking system has been used. In this model, five parallel SDD
layers of a thickness, equal to 0.5\% of radiation length each were
located at distances of 20, 46.25, 72.5, 98.75 and 125~$cm$ from the
beam-line. The SDD drift velocity, $V_{drift}$, has been assumed to be
6.75~$mm/\mu s$, which corresponds to a drift electric field of
50~$V/mm$. Five layers of the CCD based VXD of a thickness, equal to
0.12\% of radiation length each were placed at distances of 1.2, 2.4,
3.6, 4.8 and 6~$cm$. For more {\em SD\/} setup details see
Ref.~\cite{schumm:01}.

   Particle trajectories in the {\em SD\/} uniform, 5~$T$ strong
magnetic field have been generated, taking into account
small angle multiple scattering in the detector material, including
the beam pipe, VXD cryostat, air, etc. Ionization and radiative energy
losses have been ignored. The positions of the track crossing points
in each VXD and MT layer have been randomly Gaussian smeared with the
expected position resolutions: 5~$\mu m$\, for the VXD along the both
axes; 7~$\mu m$\, for the SDDs along anodes; and 10~$\mu m$\, for the
SDDs along ionization drift. Tracks were generated within the
kinematics region of $\mid\cos\theta\mid <$~0.8, where $\theta$\, is
the polar angle. The event vertex position, if used, has been assumed
to be known to $\sigma_{vtx}=$~2~$\mu m$\, for both, transverse and
longitudinal, directions. 

   In the least squares method of the helical trajectory
reconstruction, the full initial covariance matrix has been used,
taking into account the cross-correlations of the track crossing
points in the VXD and MT layers due to multiple scattering.

\section{Simulation results}

\subsection{Discrimination of tracks from event pile-up}

   In these simulations, we assume that, for the charged particles
produced in the triggered event, the event time and positions of track
crossing points in the MT layers were measured correctly. Then, at the
track reconstruction stage, the hit set, created by each single
\begin{figure}[htb]
\parbox{3.2in}{
\centerline{\hbox{
\includegraphics*[width=3.17in,bb=20 150 535 665]{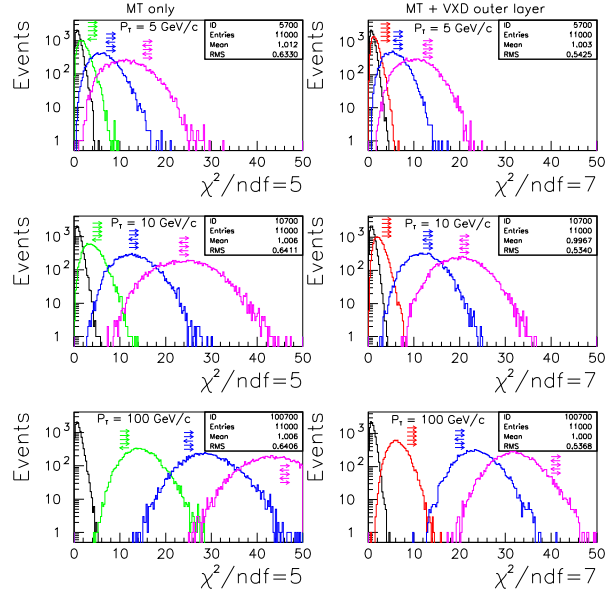}
}}
} \
\parbox{2.0in}{
\caption[]
{
\footnotesize
Examples of track discrimination for $\Delta t =$~10~$ns$\, for some
combinations of drift {\em directions\/} in participated SDDs. All
layers are drifting along {\em $z$-axis}. $\chi^{2}$\, for the correct
track timing are shown in black with no marking the relative drift
directions. In the left column (MT only), $\chi^{2}$-distributions for
all 5 SDDs, drifting in the same direction, do not differ from the
ones with correct timing. Histogram statistics are shown for the
correct timing.
\normalsize
}
\label{fig:chi2sep}
}
\end{figure}
particle, will match to a single track, yielding a good fit with
$\chi^{2}$\, within an expected range. For a track emerging before or
after the trigger by the time shift of $\Delta t$, the assumption of
its belonging to the triggered event will lead to a misplacement of
its layer crossing points by \mbox{$\pm V_{drift}\times\Delta t$}\,
along the drift axis. In the cases of at least one hit in some layer
drifting in the opposite direction compared to the others, the hit set
will not match to a single track and, thus the hypothesis that this
potential track belongs to the triggered event would be rejected. As
an illustration, in Fig.~\ref{fig:chi2sep} the simulated
$\chi^{2}$-distributions for \mbox{$\Delta t =$ 10 $ns$} are compared
to the ones with correct track timing for some combination of drift
{\em directions} in the SDD layers, all drifting along
$z$-axis\footnote{i.e. along detector magnetic field; here we switched
to the detector global coordinate system as shown on the left side of
Fig.~\ref{fig:setup}.}.

   In the 5-layer MT with all SDDs drifting along the same {\em axis},
a particle may encounter 16, roughly equally probable, distinct
combinations of the relative drift {\em directions\/} in the MT
layers. In 15 of these combinations, at least one hit drifts in the
opposite direction than the others. This means that, in
$\simeq$15/16~=~93.75\% cases, the selection procedure above will
work. Only the relatively small fraction, $\simeq$1/16~=~6.25\% of
tracks will cross the SDDs, drifting in the same direction, and the
respective out-of-time hits will still match to a single
track, but shifted as a whole from the true particle trajectory by
\mbox{$\pm V_{drift}\times\Delta t$}. In many cases, this shift can be
detected by matching the track to the VXD and/or the collision region,
and we can decide if the track should be associated with the triggered
event (see right column of Fig.~\ref{fig:chi2sep}).

   A way to avoid the necessity of using measurements beyond the MT
for sorting out all tracks in the pile-up is to let SDDs in different
layers drift along different {\em axes}: some drifting along the
\begin{figure}[htb]
\parbox{2.0in}{
\caption[]
{
\footnotesize
Examples of track discrimination for $\Delta t =$~10~$ns$\, for some
choices of drift {\em axes\/} in the MT layers and for the equally
probable combinations of drift {\em directions\/} in the crossed
SDDs. The notations like $z\varphi zzz$\, are for the particular
choices of the drift {\em axes\/} in the MT layers, starting from the
innermost one. Histogram statistics are shown for the correct timing.
\normalsize
}
\label{fig:chi2comb}
} \
\parbox{3.2in}{
\centerline{\hbox{
\includegraphics*[width=3.17in,bb=20 150 535 665]{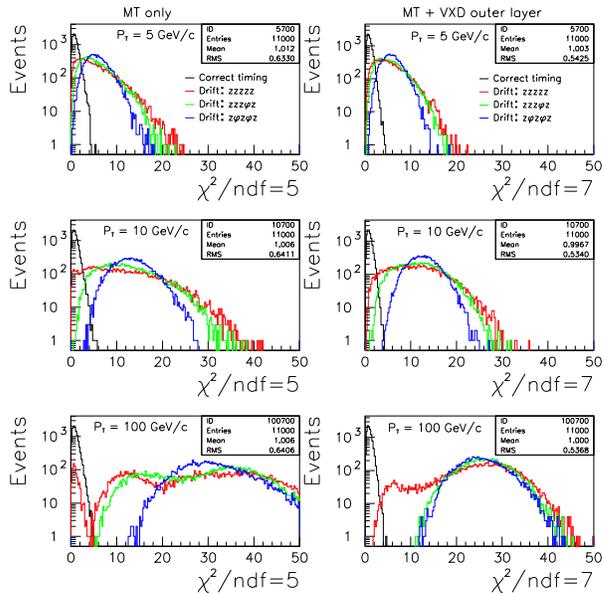}
}}
}
\end{figure}
detector magnetic field ($z$-axis) as in AGS E896~\cite{e896:00}, but
others drifting along the azimuth as in STAR at RHIC~\cite{svt:00},
i.e. along $\varphi$-axis. In Fig.~\ref{fig:chi2comb}, some choices
for the drift {\em axes} in MT layers are compared to each
other. Apparently, the best track discrimination by their generation
time is achieved by alternating drift {\em axes} from layer to layer,
like $z\varphi z\varphi z$\, (or $\varphi z\varphi z\varphi$, which is
similar but is not shown in Fig~\ref{fig:chi2comb}). But even the
options with only one layer drifting perpendicularly to the other
four provide much better results, particularly for the ``MT only'',
compared to the case of all layers drifting along the same
{\em axis}\footnote{We do not consider here another possible solution
with all SDDs drifting along the same {\em axis}, but with
significantly different drift velocities in different layers.}.

\begin{figure}[htb]
\parbox{3.0in}{
\centerline{\hbox{
\includegraphics*[width=2.9in,bb=20 155 535 650]{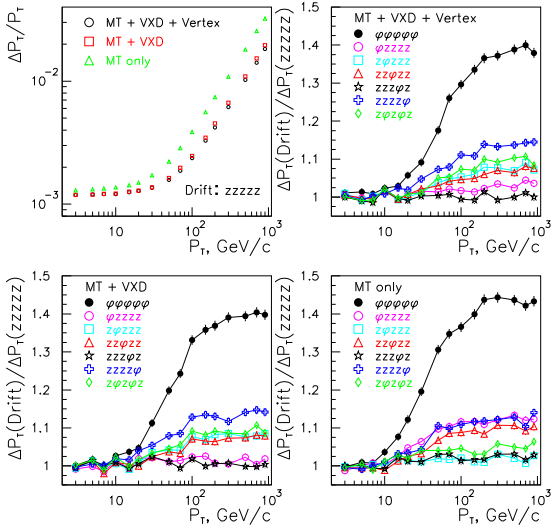}
}}
} \
\parbox{2.2in}{
\caption[]
{
\footnotesize
Momentum resolutions (RMS) for all SDDs drifting along $z$-axis (left
upper frame), and impact of various choices for the drift axes on the
{\em SD\/} tracker momentum resolution (three other frames).
\normalsize
}
\label{fig:ptres}
}
\end{figure}

   The advantages of choosing some MT layers to drift along the
azimuth should be weighted against a potential worsening of the
momentum resolution due to the expected differences in the SDD spatial
resolutions along anode and drift axes. In Fig.~\ref{fig:ptres}, the
various choices for the drift axes in the MT layers are compared with
respect to the momentum resolution. For the most important cases of
\mbox{``MT+VXD+Vertex''} and \mbox{``MT+VXD''}, the best drift
combination for the track discrimination, $z\varphi z\varphi z$, would
lead to a loss in $\Delta P_{T}/P_{T}$\, at the highest momenta by
$\sim$10\% compared to the best achievable resolution with
$zzzzz$-drift. However, the combinations $\varphi zzzz$\, and
$zzz\varphi z$ affect the momentum resolution by less than $\sim$2\%,
if at all. For the ``MT only'', the choice of $zzz\varphi z$\, would
also be one of the two best. The other combination is $z\varphi 
zzz$, but it is not among the best for the momentum resolution, using
the VXD and/or vertex. As a result, at least two choices for the drift
axes in the MT layers, $zzz\varphi z$\, and $\varphi zzzz$, should
seriously be evaluated as a good compromise between momentum
resolution and track time stamping.

\subsection{Track timing}

\begin{figure}[htb]
\parbox{2.2in}{
\caption[]
{
\footnotesize
Examples of simulated $t_{rec}-t_{0}$\, distributions for some choices
of drift {\em axes} in the MT layers and with equally probable all
combinations for drift {\em directions\/} in the participated
SDDs. Histogram statistics are shown for the $zzz\varphi z$\, drift
{\em axis\/} combination.
\normalsize
}
\label{fig:dt_histo}
} \
\parbox{3.0in}{
\centerline{\hbox{
\includegraphics*[width=2.9in,bb=20 150 540 665]{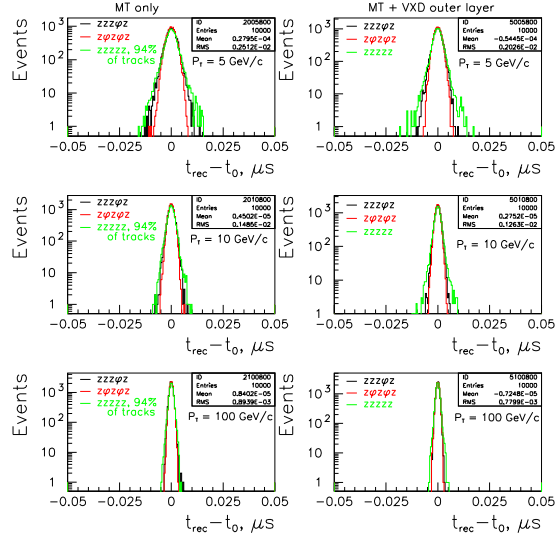}
}}
}
\end{figure}
   An additional parameter, {\em track generation time}, can be
introduced in the track fits. With this parameter derived from the
fit, each track could be assigned to an appropriate event with the
well known generation time, which has been accurately measured, using
the dedicated fast sub-detector(s). Fig.~\ref{fig:dt_histo} shows the
simulated distributions for the differences of the reconstructed,
$t_{rec}$, and actual track generation time, $t_{0}$, for three
choices of drift {\em axes} in the MT layers. One observes that the
widths of the distributions (RMS) are on the nanosecond scale, but for
high-$P_{T}$\, tracks, time resolution in the sub-nanosecond range
seems to be achievable.

\begin{figure}[htb]
\centerline{\hbox{
\includegraphics*[width=5.4in,bb=20 405 540 650]{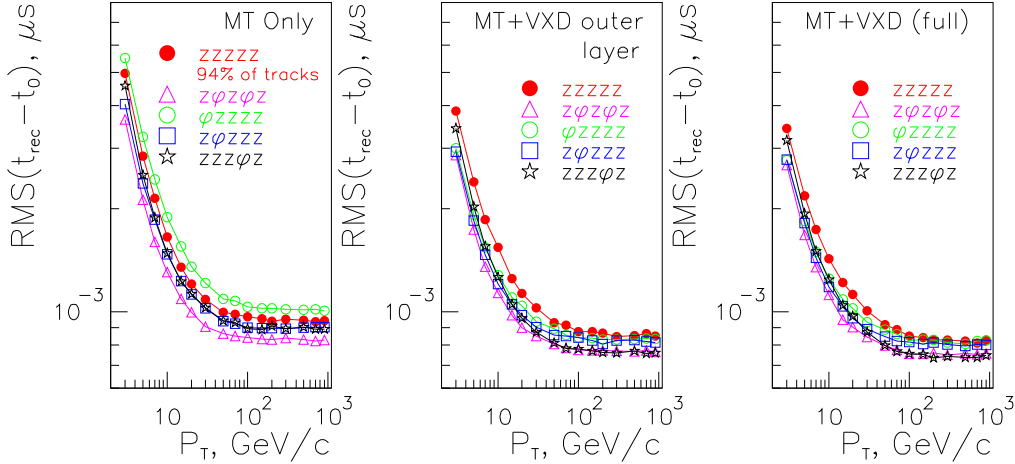}
}}
\caption[]
{
\footnotesize
Track timing (RMS) with various choices for drift axes in the MT
layers.
\normalsize
}
\label{fig:dt_of_pt}
\end{figure}

   The most promising choices for the drift {\em axes\/} in the
layers are compared in Fig.~\ref{fig:dt_of_pt}. Apparently, the
combination $z\varphi z\varphi z$\, is always the best. However, the
choice $zzz\varphi z$\, with the only one layer, drifting along the
azimuth, is hardly distinguishable from $z\varphi z\varphi z$, if the
VXD is used, and just slightly worse for the MT only case. Combined
with the earlier observation of virtually no negative impact of the
$zzz\varphi z$\, combination on the momentum resolution, this option
should be very seriously considered for the design of the SDD based
central MT.

\section{Conclusion}

   We have shown that, with an SDD based central MT for the detector
at the $e^{+}e^{-}$\, linear collider, the track selection and timing
is possible at the nanosecond and even sub-nanosecond level. This
means that, even at the NLC and/or JLC with the bunch spacing at
1.4~$ns$, each high-$P_{T}$\, track can be assigned to a particular
bunch crossing at a confidence level of up to two standard deviations.

   In order to achieve a good track timing and a minimal effect on the
momentum resolution in the proposed 5-layer central MT, we suggest a
design with four layers drifting along the magnetic field ($z$-axis),
and one layer  drifting along the azimuth ($\varphi$-axis) with
virtually no negative impact of such a choice on the detector momentum
resolution.

   It is worth underlining that with the SDD based MT, the track
timing capabilities on the sub-nanosecond scale are built into the
technology choice at no additional cost or effort.

\section{Acknowledgments}

It is our pleasure to thank K.~Riles and B.~Schumm for the useful
communications on the $e^{+}e^{-}$\, linear collider tracking system
characteristics and design, and J.~Balewski, G.~Bunce and
A.~I.~Pavlinov for the inspiring discussions. This work has been
supported in part by the DoE and NSF.



\end{document}